\documentclass[manuscript]{aastex}

\usepackage{amsmath,amssymb}

\newcommand{\bfB}{\mbox{\boldmath$B$}}
\newcommand{\bfbarB}{\overline{\mbox{\boldmath$B$}}}
\newcommand{\bfbarU}{\overline{\mbox{\boldmath$U$}}}

\newcommand{\bfg}{\mbox{\boldmath$g$}}

\newcommand{\bfk}{\mbox{\boldmath$k$}}

\newcommand\calE{{\cal E}}
\newcommand\bfcalE{\boldsymbol{\cal E}}
\newcommand\barcalE{\overline{\cal E}}
\newcommand\bfbarcalE{\overline{\boldsymbol{\cal E}}}

\newcommand{\bfalpha}{\boldsymbol{\alpha}}

\newcommand{\bfOmega}{\boldsymbol{\Omega}}
\newcommand{\bfbeta}{\boldsymbol{\beta}}
\newcommand{\Bbar}{{\overline B}}
\newcommand{\Tbar}{{\overline T}}
\newcommand{\bfb}{\mbox{\boldmath$b$}}

\newcommand{\bfe}{\mbox{\boldmath$e$}}
\newcommand{\bfu}{\mbox{\boldmath$u$}}
\newcommand{\bfU}{\mbox{\boldmath$U$}}

\newcommand{\bfnabla}{\mbox{\boldmath$\nabla$}}
\shorttitle{Mean emf driven by magnetic buoyancy instability}
\shortauthors{Davies \& Hughes}

\begin{document}

\title{The mean electromotive force resulting from magnetic buoyancy instability}

\author{C. R. Davies and D. W. Hughes}
\affil{Department of Applied Mathematics, \\ University of Leeds, Leeds LS2 9JT, U.K.}
\email{tina@maths.leeds.ac.uk, d.w.hughes@leeds.ac.uk}

\begin{abstract}
Motivated both by considerations of the generation of large-scale astrophysical magnetic fields and by potential problems with mean magnetic field generation by turbulent convection, we investigate the mean electromotive force (emf) resulting from the magnetic buoyancy instability of a rotating layer of stratified magnetic field, considering both unidirectional and sheared fields. We discuss why the traditional decomposition into $\alpha$ and $\beta$ effects is inappropriate in this case, and that it is only consideration of the entire mean emf that is meaningful. By considering a weighted average of the unstable linear eigenmodes, and averaging over the horizontal plane, we obtain depth-dependent emfs. For the simplified case of isothermal, ideal MHD we are able to obtain an analytic expression for the emf; more generally the emf has to be determined numerically. We calculate how the emf depends on the various parameters of the problem, particularly the rotation rate and the latitude of the magnetic layer.
\end{abstract}


\section{Introduction}
\label{sec:intr}

One of the most important theoretical problems in astrophysical fluid dynamics is to explain the generation of global scale magnetic fields, as observed in stars, accretion discs and galaxies. It is generally accepted that these are the result of some sort of hydromagnetic dynamo, in which the inductive motions of a highly electrically conducting plasma maintain magnetic fields against their tendency otherwise to decay. However, a convincing theoretical explanation of how such \textit{large-scale} magnetic fields can be generated --- i.e.\ fields with a significant component on scales much larger than that of the plasma motions responsible for their generation --- remains elusive.

The traditional theoretical approach to explaining the generation of large-scale magnetic fields is via mean field electrodynamics \citep[see, for example,][]{Moffatt78,KR80}, an elegant theory of magnetohydrodynamic (MHD) turbulence in which the evolution of the mean (large-scale) field is governed by a mean induction equation of the form
\begin{equation}
\frac{\partial \bfB_0}{\partial t} = \nabla \times \left( \bfU_0 \times \bfB_0 \right) + 
\nabla \times \bfcalE + \eta \nabla^2 \bfB_0 ,
\label{eq:mean_ind}
\end{equation}
where $\bfB_0$ represents the mean magnetic field, $\bfU_0$ the mean velocity, $\bfcalE$ the mean electromotive force (emf) and $\eta$ the magnetic diffusivity. The term describing the mean emf, which is the distinctive feature of equation~(\ref{eq:mean_ind}) in comparison with the unaveraged induction equation, is defined by
\begin{equation}
\bfcalE = \langle \bfu \times \bfb \rangle ,
\label{eq:mean_emf}
\end{equation}
where $\bfu$ and $\bfb$ represent the (small-scale) fluctuating velocity and magnetic fields, and angle brackets denote a spatial average over intermediate scales. The closure of equation~(\ref{eq:mean_ind}) is usually brought about by postulating an expansion of $\bfcalE$ in terms of $\bfB_0$ and its spatial derivatives,
\begin{equation}
\calE_i = \alpha_{ij}B_{0j} + \beta_{ijk} \frac{\partial B_{0j}}{\partial x_k} + \cdots \ , 
\label{eq:emf_expansion}
\end{equation}
where $\bfalpha$ and $\bfbeta$ are pseudo-tensors; \citet{HP10} discuss a more general expansion procedure involving also temporal derivatives of the mean field. In the kinematic regime, in which the field is assumed to exert no back-reaction on the flow, the components $\alpha_{ij}$ and $\beta_{ijk}$ depend solely on the properties of the velocity field and on the magnetic diffusivity. The symmetric part of the $\bfalpha$ tensor (the so-called `$\alpha$-effect') leads to field amplification, and can be non-zero only in flows that lack reflectional symmetry, such as helical flows; in its simplest isotropic form, in which $\beta_{ijk} = \beta \epsilon_{ijk}$, the scalar $\beta$ can be identified as a turbulent diffusivity, although in general it has a much more complicated interpretation \citep[see][]{KR80}. Often, astrophysical magnetic fields are modelled by assuming plausible (although somewhat arbitrary) functional forms for $\bfalpha$, $\bfbeta$ and $\bfU_0$, including the parameterisation of nonlinear effects. Via this procedure, it is possible to reproduce many of the features of observed magnetic fields, such as the solar cycle.

However, notwithstanding the ability of mean field models to be able to mimic cosmic magnetic fields, there remain some potentially serious difficulties regarding the application of the standard formulation of mean field electrodynamics to turbulent flows at high magnetic Reynolds numbers, the case of astrophysical relevance. Of particular significance for the type of model we shall consider in this paper are those that may be encountered in the kinematic regime. When the magnetic Reynolds number $Rm$ is small, everything is fine; magnetic fluctuations arise only from the interaction between the mean field and the flow, $\bfb$ is then linearly related to the mean field $\bfB_0$, the expansion~(\ref{eq:emf_expansion}) is well-founded, and equation~(\ref{eq:mean_ind}) provides the correct description of large-scale magnetic fields. However, under the astrophysical conditions of high $Rm$ turbulence, there is the likelihood that small-scale magnetic fields can be generated independently of any large-scale mean field --- \textit{small-scale dynamo action} \citep[see, for example,][]{CG95}. In this case there will be a competition between two distinct dynamo modes. The physical principles underlying large- and small-scale dynamo action are very different: the former requires a lack of reflectional symmetry in the velocity (as manifested, for example, by the flow helicity) and has a dynamo growth rate that scales with the (small) wave number of the large-scale field; the latter has a growth rate that can be related to the chaotic properties of the flow, characterised by Lyapunov exponents and cancellation exponents \citep{DO93}. The nature of the field that would be observed in this case remains a controversial issue. \citet{BCR05} and \citet{CH09} have argued that at high $Rm$, when the turbulence is rotationally influenced but not rotationally dominant, and in the absence of a large-scale velocity shear, the small-scale mode will dominate; the observed fields will then bear no relation to those predicted from any calculation of $\bfalpha$ and $\bfbeta$ and subsequent solution of equation~(\ref{eq:mean_ind}). Numerical evidence for this point of view has been provided by simulations of rotating convective turbulence \citep{CH06,HC08} and forced helical turbulence \citep{LHT07}. Conversely, if convection is strongly constrained by rotation then large-scale field generation may be possible; this was first examined by \citet{CS72}, with subsequent numerical investigations by, for example, \citet{JR00}, \citet{SH04}, \citet{KKB09}. Whether the mean field idea holds at high $Rm$ in the presence of a large-scale velocity shear remains a fascinating, but unresolved issue. Numerical simulations at moderate values of $Rm$ \citep[see, for example,][]{Yousef08,KKB08,HP09} have shown the beneficial role of a large-scale shear flow for large-scale field generation, but were not definitive in pinning down the underlying physical mechanism. \citet{PH11} have recently addressed this problem systematically by considering the nature of the magnetic fields generated by spatially-filtered flows driven by rotating convection. Intriguingly, the model offers support both to the standard mean field $\alpha \omega$ picture as well as to the idea that large-scale field generation may come from large-scale flow components (a `small scale' dynamo on the large scales). Further work is taking place to understand this extremely important issue.

Thus, given possible difficulties with the traditional kinematic $\alpha$-effect picture, together with the intrinsic interest in alternative dynamo mechanisms, we explore in this paper the conceptually very different idea of an emf resulting from an instability of the magnetic field itself. Unlike the standard picture of a weak field being amplified by a velocity field that exists in the absence of a magnetic field --- e.g.\ turbulence driven by thermal convection --- here a pre-existing magnetic field is absolutely pivotal to the generation of an emf that can subsequently lead to field amplification. Such a means of generating an emf has been well-studied in plasma confinement devices, particularly reversed field pinches \citep[see, for example,][]{Capello04}. It should though be stressed, since we are envisaging a dynamo process, that the idea of a magnetic field-driven emf does not of course provide `something for nothing'. In reversed field pinches a strong toroidal field results from external coils. In any proposed astrophysical dynamo, it is envisaged that toroidal field is wound up from poloidal field by the differential rotation of the body, and that an instability of this field can then give rise to an emf that will close the dynamo loop by regenerating the poloidal field; the underlying energy source is thus the differential rotation, which itself may be a consequence of rotating thermal convection.

Our work is particularly motivated by considerations of a possible dynamo located in the solar tachocline, with the idea that the strong velocity shear could generate toroidal field from a weak poloidal component, and in which the poloidal field could be maintained via an emf driven by magnetic buoyancy instability, the mechanism believed to be responsible for the escape of magnetic flux from the deep interior. The basic idea of magnetic buoyancy instability is that a horizontal magnetic field that decreases with height will `puff up' the atmosphere, and that if the field gradient is sufficiently strong then instability can ensue. In its details though it can be quite a subtle instability, with 3D modes preferred, despite their having to overcome the stabilising effects of magnetic tension, and with further complications arising from the roles of rotation and diffusion. A recent review of magnetic buoyancy instabilities can be found in \citet{Hughes07}.

At the onset of instability, the fastest growing linear eigenfunction will dominate, and hence, at least in the early stages of the instability, the ensuing emf will be driven by this linear mode. In this paper we therefore investigate the nature of the emf that results from the rotationally influenced magnetic buoyancy instability.  We consider the instability in fairly general terms, retaining all possible modes of instability, and considering two types of horizontal equilibrium magnetic fields: unidirectional fields of the form $\bfB = (B_x(z),0,0)$, and what in plasma physics are often referred to as `sheared' fields, taking the form $\bfB = (B_x(z),B_y(z),0)$. We therefore build upon the work of \citet{Moffatt78} and \citet{Schmitt84, Schmitt03} who considered a similar problem, but restricted attention to magnetostrophic instabilities of a unidirectional field. \citet{Thelen2000} also calculated the emf resulting from a magnetic buoyancy instability, but considered the somewhat different case of Rayleigh-Taylor type instabilities driven by the density discontinuity at the upper interface of a layer of field embedded in a non-magnetic atmosphere. An ostensibly similar problem, though actually one with rather different physical foundations, has been considered by \citet{FMSS94}, who calculated the emf arising from the instability of isolated magnetic flux tubes; the differences between the instabilities of a continuously stratified field and those of isolated flux tubes, which are significant, are discussed in \citet{Hughes07}. In an interesting model with a different set-up, \cite{CBC03} have shown, via fully nonlinear numerical simulations, how dynamo action may be achieved by the combined action of a velocity shear and magnetic buoyancy; for their model, in contrast to that discussed in this paper, the flow and the instability are on the same scale, so a mean field picture is not appropriate.

The notion of an instability-driven dynamo has also been discussed with regard to other instability mechanisms, although not in terms of generating a mean electromotive force that can close the dynamo loop. The primary candidate for causing fluid turbulence in accretion disks, necessary for angular momentum transfer, is the magnetorotational instability (MRI), which, via only a weak magnetic field, can lead to the instability of a differentially rotating disk that is hydrodynamically stable \citep{BH91}. By a `bootstrapping' mechanism, the resulting three-dimensional velocity field can then act to amplify the magnetic field that, via its instability, had created the velocity field originally \citep{BH98}. In terms of magnetic field generation, the role of the MRI is thus to transform the velocity field from a state of purely differential rotation, which cannot act as a dynamo, to a three-dimensional turbulent state, which can. However, exactly how this mechanism may work remains poorly understood. Many early numerical simulations produced healthy-looking dynamos, but often these employed only numerical dissipation. The dangers of such an approach have been highlighted by \citet{FP07}, and it has become clear that if an understanding is to be obtained computationally, high-resolution simulations with explicit dissipation will be needed.

\citet{Spruit02} has discussed, in general terms, how a dynamo might operate in convectively stable regions of stellar interiors, driven by differential rotation and a Tayler instability of the magnetic field (rather than magnetic buoyancy, as discussed here). Order of magnitude estimates are given of the strength of the magnetic field that may result from such a mechanism, should it work. However, as pointed out by \citet{MZB08}, the closing of the dynamo loop in Spruit's model comes directly from the small-scale poloidal field, and not, as it should, from the large-scale emf resulting from the instability. This is an important point, since an instability \textit{per se} is not enough, as can be seen by consideration of two-dimensional instabilities; it is the mean emf that is crucial. The numerical simulations of \citet{MZB08} reproduce the Tayler instability, but, in the parameter regimes they explored, they do not find dynamo action.

The underlying principles of instability-driven dynamos are very different from those of `traditional' dynamo mechanisms. The latter are characterised by the exponential amplification of a weak seed magnetic field, which grows until the nonlinear influence of the Lorentz force back on the flow allows equilibration to a statistically steady state. The evolution of the field can be split into distinct kinematic and dynamic regimes; for the former, it makes sense to discuss properties of the velocity field itself that may be beneficial for magnetic field generation. Furthermore, for kinematic mean field dynamos one can decompose the mean electromotive force into distinct components proportional to the large-scale field and to its gradient, as suggested by equation~(\ref{eq:emf_expansion}), although the caveats discussed above regarding the difficulties of this approach at high values of $Rm$ need to be borne in mind. For instability-driven dynamos, the whole picture is very different since the perturbations to the flow and magnetic field needed to drive the emf come about only through an initial instability of the field itself. There is therefore no concept of a strictly kinematic regime in which the flow can be described independently of the magnetic field. Furthermore, although, as we shall show below, it is possible to describe such dynamos within the mean field framework, it is in general \textit{not} meaningful to speak of $\alpha$ and $\beta$-effects. The instability, and hence the resulting emf, will typically have a complicated nonlinear dependence on the strength and gradients of the magnetic field, rendering meaningless a decomposition such as (\ref{eq:emf_expansion}). An exception to this might be the magnetorotational instability, which is a weak field instability. In this case, \citet{Bburg05} and \citet{Gressel10} have argued that a mean field expansion of the form~(\ref{eq:emf_expansion}) is possible, although it should be stressed, as acknowledged by \citet{Gressel10}, that a more symmetric treatment that treats the momentum and induction equations on an equal footing, an approach pioneered by \citet{CHP10a,CHP10b}, is required.

Obtaining a mathematical description of an instability-driven dynamo is less straightforward than for a standard, flow-driven dynamo, where, at least in the kinematic regime, the problem is tackled either via the induction equation or the mean induction equation. For an instability-driven dynamo, the starting point is an analysis of the pertinent instability. The difficulty then arises in describing how this instability feeds back into the dynamo process. The initial phases of the instability will be characterised by exponential growth, and it therefore makes sense as a first step to consider the nature of the emf (which will also be growing exponentially) resulting from the most unstable mode of instability and its dependence on the various parameters of the problem. The ultimate aim must be to try to incorporate, self-consistently, the emf arising from the instability, in both its linear and nonlinear phases, with the generation of toroidal field from differential rotation.

Our aim in this paper is to evaluate the nature of the emf arising from magnetic buoyancy instability of a rotating, stratified layer of gas containing a horizontal, depth-dependent, but not necessarily unidirectional, magnetic field. In \S\,\ref{sec:formulation} we describe the formulation of the instability problem and explain the prescription we use for calculating the resulting emf. In \S\,\ref{sec:ideal} we consider the instability of a unidirectional field for the simplified case of isothermal, ideal MHD; this allows an analytical solution for the emf, and thus highlights its complex dependence on the magnetic field and its gradient. This also allows us to discuss the nature of the emf in the case of perfect electrical conductivity, an issue that has given rise to problems in the standard mean field approach \citep{Moffatt78}. Section~\ref{sec:unidirectional} considers the more general instability analysis, with all the diffusivities restored, which has to be performed numerically. In Section~\ref{sec:sheared} we extend our analysis to consider the emf resulting from the instability of a sheared magnetic field. Our conclusions and a discussion of how the work may be extended are contained in \S\,\ref{sec:conc}.

\section{Formulation of the Instability Problem}

\label{sec:formulation}

We consider a plane layer of fluid of depth $d$ located at colatitude $\theta$, rotating with angular speed $\Omega$ about the $\theta = 0$ axis. We employ Cartesian coordinates, with $x$, $y$ and $z$ facing east, south and inwards respectively. The top of the layer is assumed to be the $z=0$ plane. The angular velocity is given by $\bfOmega = \Omega(0, -\sin\theta, -\cos\theta) = \Omega \bfk$. For simplicity, the fluid is assumed to be isothermal, with infinite thermal diffusivity.

In standard notation, the governing equations of compressible MHD in a rotating frame may be expressed as
\begin{equation}
\rho \frac{\mbox{D} \bfu} {\mbox{D} t} + 
2 \rho (\mathbf{\Omega} \times \bfu ) = 
- \mathbf{\nabla}p + 
\frac{1}{\mu_{0}}(\mathbf{\nabla} \times \bfB ) \times \bfB +
\rho g \bfe_z +
\mu \left ( \mathbf{\nabla}^{2} \bfu + \frac{1}{3} \mathbf{\nabla}(\mathbf{\nabla} \cdot \bfu ) \right ) ,
\label{eq:momentum}
\end{equation}
\begin{equation}
\frac{\partial \bfB }{\partial t} = \nabla \times (\bfu \times \bfB ) + \eta \mathbf{\nabla}^2 \bfB ,
\label{eq:induction}
\end{equation}
\begin{equation}
\frac{\mbox{D} \rho}{\mbox{D} t} + \rho\mathbf{\nabla} \cdot \bfu = 0 ,
\label{eq:continuity}
\end{equation}
\begin{equation}
\nabla \cdot \bfB = 0 ,
\label{eq:divB}
\end{equation}
\begin{equation}
p = \rho c^2 \mbox{,}
\label{eq:state}
\end{equation} 
where $c=(\mathcal{R}\Tbar)^{1/2}$ is the isothermal sound speed, $\eta$ is the magnetic diffusivity (assumed constant), $\mu$ is the viscosity (also assumed constant) and $\mu_0$ is the magnetic permeability.

We consider the instability of stationary equilibrium states, with uniform horizontal magnetic fields, linearly dependent on depth. In the stability analyses of \S\S\,\ref{sec:ideal},\,\ref{sec:unidirectional} the equilibrium field is assumed to be of the form $\bfB = B_0(1+\zeta z /d)\bfe_x$, which we regard as the toroidal field. However, although it is thought that the toroidal field in the solar interior is dominant, it is of interest to consider the role of an additional weak poloidal ingredient; in \S\,\ref{sec:sheared} we therefore extend our analysis to include equilibrium fields of the form $\bfB = B_0(1+\zeta z/d)\bfe_x + B_1(1+\xi z/d)\bfe_y$.

Using equations~(\ref{eq:momentum}) -- (\ref{eq:state}), and eliminating the pressure, linear perturbations to the basic state are governed by the following equations, here expressed in dimensionless form:
\begin{align}
\nonumber
\bar{\rho} \frac{\partial \bfu }{\partial t} + 2 \tilde \Omega \bar{\rho} \left( \bfk \times \bfu \right) = 
- \mathbf{\nabla} \rho &+
A^2 \left( (\bfb \cdot \nabla) \bfbarB + (\bfbarB \cdot \nabla) \bfb - \nabla (\bfbarB \cdot \bfb ) \right) 
+ \chi \rho \bfe_z \\
& \qquad \qquad \qquad \qquad
+ \tilde \nu \left( \mathbf{\nabla}^2 \bfu + \frac{1}{3} \nabla (\nabla \cdot \bfu ) \right) ,
\label{eq:momentum_pert}
\end{align}
\begin{equation}
\frac{\partial \bfb}{\partial t} = \nabla \times \left( \bfu \times \bfbarB \right) + \tilde \eta \nabla^2 \bfb ,
\label{eq:induction_pert}
\end{equation}
\begin{equation}
 \frac{\partial \rho}{\partial t} + \nabla \cdot \left( \bar{\rho} \bfu \right) = 0 ,
\label{eq:continuity_pert}
\end{equation}
\begin{equation}
\nabla \cdot \bfb = 0 \mbox{,}
\label{eq:divb}
\end{equation}
where barred variables represent equilibrium values and unbarred variables represent perturbations. The equations are scaled using velocity $c$ (the isothermal sound speed), length $d$ (the layer depth), density $\rho_{0}=\bar{\rho}(z=0)$ and magnetic field strength $\overline{B}_{x0} = \overline{B}_{x}(z=0)$. The dimensionless parameters in equations~(\ref{eq:momentum_pert}) -- (\ref{eq:divb}) are defined by
\begin{equation}
\tilde \Omega = \frac{\Omega d}{c} \, , \qquad
A^2 = \frac{B_0^2 / \mu_0 \rho_0}{c^2} \, , \qquad
\chi = \frac{gd}{c^2}  \, , \qquad
\tilde \nu = \frac{\mu}{\rho_{0}cd} \, , \qquad 
\tilde \eta = \frac{\eta}{cd} \, .
\label{eq:dim_constants}
\end{equation}
From now on we shall work only with dimensionless parameters, so, for simplicity, we shall drop the tildes.

The equilibrium density distribution is calculated from the $z$-component of equation~(\ref{eq:momentum}), after making use of (\ref{eq:state}); in dimensionless form this may be written as 
\begin{equation}
\frac{\partial \bar{\rho}}{\partial z} - \chi \bar{\rho} =
- \frac{A^2}{2} \frac{\partial}{\partial z} \Bbar^2 \mbox{.}
\label{eq:eq_state}
\end{equation}
With the dimensionless equilibrium magnetic field written as $\bfbarB = (1+\zeta z) \bfe_x + \lambda (1+\xi z) \bfe_y$, the solution of (\ref{eq:eq_state}) is given by 
\begin{align}
\nonumber
\bar{\rho} = \left( 1-\frac{A^2}{\chi} \left( \zeta \left( 1+\frac{\zeta}{\chi} \right) + 
\lambda^2 \xi \left( 1+\frac{\xi} {\chi} \right) \right) \right) e^{\chi z} &+ 
\frac{A^2}{\chi} \left( \zeta \left( 1 + \frac{\zeta}{\chi} \right) 
+ \lambda^2 \xi\left(1+\frac{\xi}{\chi} \right) \right) \\ &+ 
\frac{A^2}{\chi}(\zeta^2 + \lambda^2 \xi^2 )z ,
\label{eq:rhobar}
\end{align}
where $\lambda=B_{1}/B_{0}$ (the ratio of the strengths of the fields in the $y$- and $x$-directions at $z=0$).

Since the equilibrium state is dependent only on $z$, the perturbations take the form $\Re (a(z)(e^{i(kx+ly)+st}))$, where $a(z)$ is an amplitude function (which may be complex). The perturbation equations then become
\begin{align}
\nonumber
su = -\left( \frac{4}{3}k^2+l^2 \right) \frac{\nu}{\bar{\rho}}u - 
\left( 2 \Omega \cos \theta + \frac{kl}{3} \frac{\nu}{\bar{\rho}} \right) v 
&+ 2 \Omega \sin \theta \, w + il \frac{A^2}{\bar{\rho}} \Bbar_y b_x 
-ik \frac{A^2}{\bar{\rho}}\Bbar_y b_y \\
&+ \frac{A^2}{\bar{\rho}}\Bbar_x^{\prime} b_z
- \frac{ik}{\bar{\rho}}\rho 
+ \frac{ik}{3}\frac{\nu}{\bar{\rho}} w^{\prime} + \frac{\nu}{\bar{\rho}}u^{\prime\prime} ,
\label{eq:u_pert}
\end{align}
\begin{align}
\nonumber
sv = \left( 2\Omega \cos \theta -\frac{kl}{3} \frac{\nu}{\bar{\rho}} \right) u &- \left( k^2 
+ \frac{4}{3}l^2 \right) \frac{\nu}{\bar{\rho}}v 
- il\frac{A^2}{\bar{\rho}}\Bbar_x b_x + ik \frac{A^2}{\bar{\rho}}\Bbar_x b_y \\
&+ \frac{A^2}{\bar{\rho}}\Bbar_y^{\prime} b_z 
- \frac{il}{\bar{\rho}}\rho + \frac{il}{3}\frac{\nu}{\bar{\rho}}w^{\prime} 
+ \frac{\nu}{\bar{\rho}}v^{\prime\prime}  ,
\label{eq:v_pert}
\end{align}
\begin{align}
\nonumber
sw = -2 \Omega \sin \theta \, u -
(k^2+l^2)\frac{\nu}{\bar{\rho}}w &- 
\frac{A^2}{\bar{\rho}}\Bbar_x^{\prime} b_x -
\frac{A^2}{\bar{\rho}}\Bbar_y^{\prime} b_y + 
\frac{A^2}{\bar{\rho}} \left( ik \Bbar_x + il \Bbar_y \right) b_z + 
\frac{\chi}{\bar{\rho}}\rho \\ 
&+ \frac{ik}{3}\frac{\nu}{\bar{\rho}} u^{\prime} +
\frac{il}{3}\frac{\nu}{\bar{\rho}} v^{\prime} - 
\frac{\rho^{\prime}}{\bar{\rho}} -
\frac{A^2}{\bar{\rho}}\Bbar_x b_x^{\prime} -
\frac{A^2}{\bar{\rho}}\Bbar_y b_y^{\prime} +
\frac{4}{3}\frac{\nu}{\bar{\rho}}w^{\prime\prime} ,
\label{eq:w_pert}
\end{align}
\begin{equation}
s b_x = il \Bbar_y u - il \Bbar_x v - \Bbar_x^{\prime} w -(k^2+l^2) \eta b_x - \Bbar_x w^{\prime} 
+ \eta b_x^{\prime\prime} , 
\label{eq:bx_pert}
\end{equation}
\begin{equation}
s b_y = -ik \Bbar_y u + ik \Bbar_x v -\Bbar_y^{\prime} w - (k^2+l^2) \eta b_y - \Bbar_y w^{\prime} + 
\eta b_y^{\prime\prime} ,
\label{eq:by_pert}
\end{equation}
\begin{equation}
s b_z = i(k\Bbar_x + l \Bbar_y )w - (k^2+l^2)\eta b_z + \eta b_z^{\prime\prime} ,
\label{eq:bz_pert}
\end{equation}
\begin{equation}
s \rho = -ik \bar{\rho}u - il \bar{\rho}v - \bar{\rho}^{\prime}w - \bar{\rho}w^{\prime}\mbox{,}
\label{eq:rho_pert}
\end{equation}
where $\bfbarB = (\Bbar_x,\Bbar_y,0)$, and where a prime denotes differentiation with respect to $z$. 

An eigenvalue problem for the growth rate $s$, which may be complex, is formulated by dividing the layer between $z=0$ and $z=1$ into $7N$ intervals (typically we take $N=600$), and using fourth order finite differences to approximate derivatives with respect to $z$. This allows the problem to be written in the form $A \mathbf{x} = s \mathbf{x}$, where $A$ is a banded matrix and where the eigenvector $\mathbf{x}$ takes the form $(\bfu_0, \bfb_0, \rho_0 , \bfu_1, \bfb_1, \rho_1, \ldots, \bfu_N, \bfb_N, \rho_N)$. The eigenvalue problem is solved using the method of inverse iteration, which converges rapidly to the eigenvalue closest to an initial guess.

The electromotive force for a particular mode of instability is determined by an average over the horizontal plane and, if the mode has a non-zero frequency, also over the period of oscillation. This leads to a $z$-dependent mean emf, analogous in form to that considered by \citet{CS72} in their study of mean field generation in a rotating convective layer. The components of $\bfcalE = \langle \bfu \times \bfb \rangle$ are given by
\begin{align}
\mathcal{E}_x &= 1/2(v_r b_{zr} + v_i b_{zi} - w_r b_{yr} - w_i b_{yi}) , \label{eq:Ex} \\ 
\mathcal{E}_y &= 1/2(w_r b_{xr} + w_i b_{xi} - u_r b_{zr} - u_i b_{zi}) , \label{eq:Ey} \\ 
\mathcal{E}_z &= 1/2(u_r b_{yr} + u_i y_{zi} - v_r b_{xr} - v_i b_{xi}) , \label{eq:Ez}
\end{align}
where $u(z)= u_r + i u_i$, etc. As in any theory in which nonlinear terms are constructed from products of terms determined from a linear stability analysis, one has to deal with the issue of the amplitude of the terms, which, of course, is in some sense arbitrary. Here we choose to normalise the eigenfunctions with the total energy (kinetic + magnetic) of the perturbation. Since it is the curl of $\bfcalE$ that is of dynamical significance, rather than $\bfcalE$ itself, then only $\mathcal{E}_x$ and $\mathcal{E}_y$ are of importance.

A crucial point to recognise is that if there is more than one mode of instability then these will have different, and possibly competing, emfs \citep{Schmitt84, Schmitt03}. If there is a unique mode of maximum growth rate then clearly the emf from this mode will dominate. If, however, there are two (or more) distinct modes of (equal) maximum growth rate then the emfs from each mode need to be considered. For example, in the non-rotating, unidirectional field case (i.e.\ when $\Bbar_y=0$ and $\Omega=0$), two modes with wave vectors $(k,l)$ and $(-k,l)$ (where $k,l>0$) have the same growth rate, but produce different emfs, with equal $\calE_y$ but equal and opposite $\calE_x$. Thus, overall, there is no contribution to $\calE_x$, as might be expected from standard mean field considerations for flows lacking reflectional symmetry (though it is important to remember that here we are dealing with the entire emf, not just the $\alpha$-effect). Similarly, symmetry considerations suggest that, in the presence of rotation, $\calE_x$ should vanish at the equator.



Thus, in order to ensure that the emf obeys the symmetries of the problem, and also that it varies smoothly with parameters (for example, that $\calE_x=0$ when $\Omega=0$ or at the equator), it is necessary to apply a weighting procedure between competing modes. Our prescription for the unidirectional field case is as follows. We determine the two modes of maximum growth rate with $l>0$, one with $k>0$ and one with $k<0$. If the two modes have growth rates $s_1$ and $s_2$, with $s_1 \geq s_2$, and emfs $\bfcalE_1$ and $\bfcalE_2$, respectively, then we define the resulting emf by
\begin{equation}
\bfcalE = \bfcalE_1 + e^{s_2/s_1} \bfcalE_2 .
\label{eq:weighted_emf}
\end{equation}
Such a prescription allows a smooth transition as $\theta$ passes through $\pi/2$, for example. At the equator, modes with equal $k$, but oppositely-signed $l$ have equal growth rates ($s_1=s_2$) and $\calE_x$ and $\calE_z$ of equal magnitude, but opposite sign. Away from the equator there is a definitely preferred mode. Expression (\ref{eq:weighted_emf}) incorporates these features. We employ a similar procedure for the instability of a sheared magnetic field, discussed in more detail in \S\,\ref{sec:sheared}.

Finally in this section, we comment on the general form of the emf that may be expected simply from considerations of the geometry of the problem. \citet{KR80} derive an expression for $\bfcalE$ in the case where there are three preferred directions, intended to represent the angular velocity $\bfOmega$, the mean velocity field $\bfbarU$ (which is zero in the present formulation) and the direction of gravitational attraction $\bfg$, with the only unknowns being several scalar quantities. Assuming a linear dependence on each of the listed vector quantities, \citet[][equation (5.45)]{KR80}, derive the following expression for the mean emf:
\begin{align}
\nonumber
\bfcalE = - \beta \nabla \times \bfbarB
- \gamma \bfg \times \bfbarB
- \gamma'' \bfbarU \times \bfbarB
- \beta_3 \left( \bfOmega \cdot \nabla \right) \bfbarB 
&- \beta_2 \nabla \left( \bfOmega \cdot \bfbarU \right)
- \alpha_1' \left( \bfg \cdot \bfOmega \right) \bfbarB \\
&- \alpha_2' \left( \bfg \cdot \bfbarB \right) \bfOmega
- \alpha_3' \left( \bfOmega \cdot \bfbarB \right) \bfg .
\end{align}
Applying this to the formulation described above gives
\begin{align}
\bfcalE &= \left( B_0 \left( \beta_3 \Omega \cos \theta \frac{\zeta}{d}
+ \alpha'_1 \Omega g \cos \theta \left( 1 + \frac{\zeta z}{d} \right) \right)
+ B_1 \left( \beta \frac{\xi}{d} + \gamma g \left( 1 + \frac{\xi z}{d} \right) \right) \right) \mathbf{e}_x 
\nonumber \\
&+ \left( B_1 \left( \beta_3 \Omega \cos \theta \frac{\xi}{d}
+\alpha'_1 \Omega g \cos \theta \left( 1 + \frac{\xi z}{d} \right) \right)
- B_0 \left( \beta \frac{\zeta}{d} + \gamma g \left(1 + \frac{\zeta z}{d} \right) \right) \right) \mathbf{e}_{y}
\label{eq:radler}\\ 
&+ \alpha'_3 \Omega \sin \theta B_1 \left( 1 + \frac{\xi z}{d} \right) g \mathbf{e}_{z} , \nonumber
\end{align}
where $\alpha_{1}^{\prime}$, $\alpha_{3}^{\prime}$, $\beta$, $\beta_{3}$ and $\gamma$ are undetermined scalars.

Owing to the influence of boundary conditions, we do not expect the emfs to have the exact form predicted by the above equations; we do however expect that the variation of the emf with the defining parameters of the problem will be similar. This can be measured by evaluating, for example, a root mean square emf defined by $\bfbarcalE = \left( \int{\bfcalE^2 \mbox{d}z} \right)^{1/2}$. The above equations can be used to predict the dependence of $\barcalE_i$ on a selection of parameters, and these results compared with those obtained by solving the full eigenvalue problem.

\section{Ideal MHD Analysis of the Electromotive Force}
\label{sec:ideal}

\subsection{Calculation of the emf}

The first calculation of the electromotive force resulting from magnetic buoyancy instability was by \citet{Moffatt78}, whose calculation was based on the stability analysis of \citet{Gilman70}, with a small number of further approximations. In the analysis below, a similar method is followed, although no additional approximations are made beyond those of \citet{Gilman70}. We consider a rotating, isothermal layer of fluid, with sound speed $c$, equilibrium magnetic field $\Bbar\mathbf{e}_{x}$ and gravity $\bfg = g \bfe_z$, and with viscous and resistive effects neglected. Following \citet{Gilman70}, we shall consider only the component of rotation perpendicular to gravity --- i.e.\ all terms involving $\cos\theta$ are neglected. As in \S\,\ref{sec:formulation}, perturbations are taken to be proportional to $\exp \left( i(kx+ly)+ st) \right)$. The (dimensional) equations governing the perturbation amplitudes as functions of $z$ (cf.\ equations~(\ref{eq:u_pert}) -- (\ref{eq:rho_pert})) can then be written as 
\begin{align}
s \bar{\rho}u &= 2 \Omega \sin \theta \bar{\rho} w - ikp + b_z \Bbar^{\prime} , \label{eq:rho_pert_Gilman} \\ 
s \bar{\rho}v &= -il(p+b_x \Bbar) + ikb_y \Bbar , \\ 
s \bar{\rho}w &= -2 \Omega \sin \theta \bar{\rho}u - p^{\prime}
- \Bbar^{\prime} b_x - \Bbar b_x^{\prime} + ik \Bbar b_z + \rho g ,\\ 
s b_x &= -il \Bbar v - \Bbar^{\prime} w - \Bbar w^{\prime} , \\ 
s b_y &= ik \Bbar v , \\
s b_z &= ik \Bbar  w ,\\
s \rho &= - ik \bar{\rho}u - il \bar{\rho} v - \bar{\rho}^{\prime} w - \bar{\rho} w^{\prime} . \\
p &= \rho c^{2} \label{eq:p_rho_Gilman} ,
\end{align}


The emf in the $x$-direction is given by $\mathcal{E}_x = 1/2\Re(vb_z^* - wb_y^*)$. Using the above equations to evaluate this gives
\begin{equation}
\calE_x = s_r \frac{k}{l} \frac{\Bbar}{|s^2|}
\left( |w|^2 \Re(Q)+ \frac{1}{2}\frac{d|w|^2}{dz} \right) ,
\label{eq:emfx_Gilman}
\end{equation}
where 
\begin{equation}
Q=\frac{s^2 (c^2 (\ln \bar{\rho})^{\prime} +
A^2 (\ln \Bbar)^{\prime}) + i 
2kc^2 s \Omega \sin \theta}{s^2 (A^2 +c^2) + k^2 A^2 c^2},
\end{equation}
$s_r= \Re(s)$ is the growth rate, and where we have used the fact that the most unstable mode has small $k$ and large $l$ \citep{Gilman70}. The eigenvalue $s$ depends on each of the variables in the basic state formulation, including the large-scale magnetic field $\bfbarB$. Thus, even in this extremely simplified case, the emf cannot be meaningfully expressed in the form of expression~(\ref{eq:emf_expansion}). It is similarly unreasonable to expect that the emfs calculated for the more complex problems considered in \S\S\,\ref{sec:unidirectional},\,\ref{sec:sheared} can be decomposed in this way.

A similar expression can be derived for $\calE_y = 1/2 \Re ( w b_x^* - u b_z^* )$. As shown by \citet{Gilman70}, $v$ and $b_y$ scale as $1/l$, leading to $\calE_x$ and $\calE_z$, but not $\calE_y$, scaling as $k/l$. Thus $\calE_y$ dominates in magnitude in this example, a feature that we may expect also to hold for the full problem incorporating diffusion.

\subsection{A well-defined emf in the absence of magnetic diffusion}

It is of interest, at this stage, to point out an important mathematical difference, for the case of a perfectly conducting fluid ($\eta =0$), between the standard kinematic formulation of the mean emf and that resulting from an MHD instability. In the traditional approach, a formal solution of the ideal (diffusionless) induction equation in terms of the Cauchy solution leads to expressions for $\alpha_{ij}(t)$ and $\beta_{ijk}(t)$ in terms of integrals along fluid particle paths of averaged quadratic Lagrangian quantities \citep{Moffatt74}. However, as pointed out by \citet{Moffatt74}, there is no guarantee that these integrals will converge as $t \rightarrow \infty$, and thus the status of the results is unclear. By contrast, for the instability-driven emf described above --- and, we believe, for instability-driven emfs in general --- there is nothing singular about the behaviour when $\eta=0$. Expression~(\ref{eq:emfx_Gilman}), calculated with $\eta=0$, is well-defined and, were we to incorporate weak magnetic diffusion, the resulting emf would simply be a regular perturbation to the result~(\ref{eq:emfx_Gilman}).

The key difference between the two systems lies in the spectra of the mathematical operators describing the two linear problems. For any non-zero value of $\eta$, however small, the induction operator possesses a discrete spectrum, with well-defined eigenfunctions; however, for $\eta=0$ the discrete spectrum disappears, together with the eigenfunctions. Indeed, this singularity in the induction operator lies at the heart of fast dynamo theory \citep[see, for example,][]{CG95}. By contrast, the linear operator for the stability problem has a discrete spectrum of unstable modes and associated eigenfunctions, leading to an unambiguous determination of the electromotive force.

\section{Unidirectional Magnetic Field}
\label{sec:unidirectional}

In this section we consider the emf resulting from the instability of a unidirectional mean magnetic field of the form $\bfbarB=B_{0}(1+\zeta z/d)\mathbf{e}_{x}$, with non-zero values of the viscosity and magnetic diffusivity. The instability evolves according to equations~(\ref{eq:u_pert})--(\ref{eq:rho_pert}) (with $\Bbar_y =0$), and the mean emf is calculated via horizontal averaging, given by (\ref{eq:Ex})--(\ref{eq:Ez}), and the weighting procedure~(\ref{eq:weighted_emf}).

\begin{figure}[htb]
\plotone{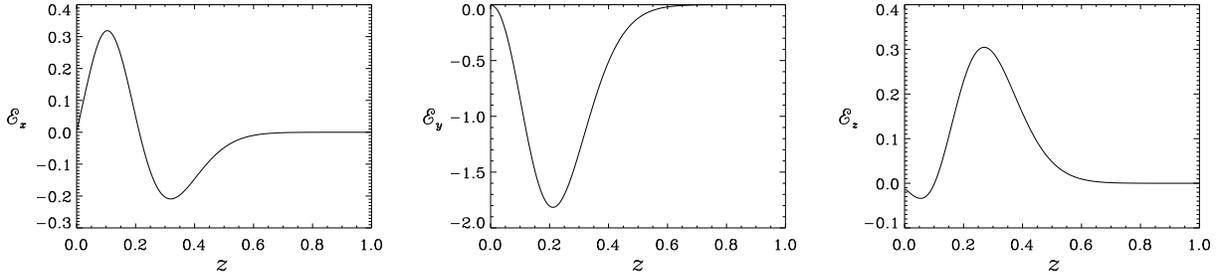}
\caption{$\calE_x$, $\calE_y$ and $\calE_z$ versus depth, with $\theta=75^{\circ}$, $\Omega=0.1$, $A^{2}=0.1$, $\zeta=1$, $\tilde \nu = \tilde \eta = 5\times 10^{-5}$, $\chi =2$.}
\label{fig:emfs}
\end{figure}

Figure~\ref{fig:emfs} shows the three components of $\bfcalE$ plotted as functions of $z$, for a test case at colatitude $\theta = 75^{\circ}$, representing the region in which most magnetic activity is observed on the solar surface. For this parameter set, the modes with largest growth rates have wave numbers $(k,l)=(0.70367,14.812)$ (with complex growth rate $s=0.082626-0.067823i$) for $k>0$, and $(k,l)=(-0.70334,14.812)$ (with $s=0.082602+0.067830i$) for $k<0$. The emfs in the figure are calculated from the mode with largest growth rate; when there is a unique mode of maximum growth rate, as here, the differences between the weighted and unweighted emfs are small.


\begin{figure}[htb]
\plotone{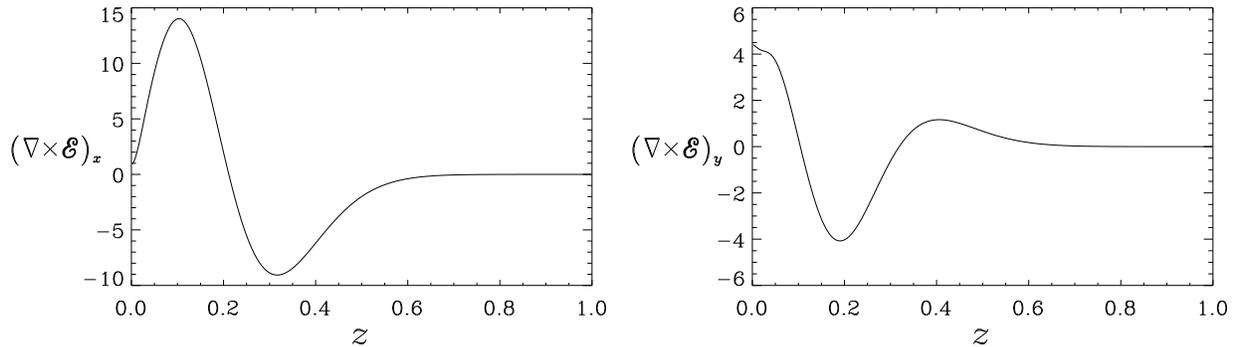}
\caption{The $x$- and $y$-components of $\bfnabla\times\bfcalE$ for the same parameter values as in Fig.~\ref{fig:emfs}.}
\label{fig:curle}
\end{figure}

It is of interest to note that the $y$-component of the emf (i.e.\ perpendicular to the imposed magnetic field) has the largest magnitude. The calculations in \S\,\ref{sec:ideal} indicate that in the absence of viscosity and magnetic diffusivity, the ratio $\calE_x/\calE_y$ is $O(k/l)$, which is vanishingly small for the preferred mode of $l \to \infty$. Although now all diffusivities are restored, the physical mechanism underlying the instability is unchanged, so we may still expect the ratio of the emfs to be $O(k/l)$, and hence small. It can also be seen that the emf is localised towards the top of the layer. Consideration of the simplest instability criterion given by \citet{Gilman70}, showing that a layer is unstable if the magnetic field strength decreases with height, yields for the current problem the condition $\zeta/(1+\zeta z)>0$. The left-hand side of this inequality is clearly largest at the top of the layer, leading to the eigenfunctions (and hence the emfs) being peaked in this region. Figure~\ref{fig:curle} shows the $x$- and $y$-components of $\bfnabla\times\bfcalE$; these are the dynamically significant elements in the evolution of the mean field, as explained in \S\,\ref{sec:formulation}.

\begin{figure}[htb]
\plotone{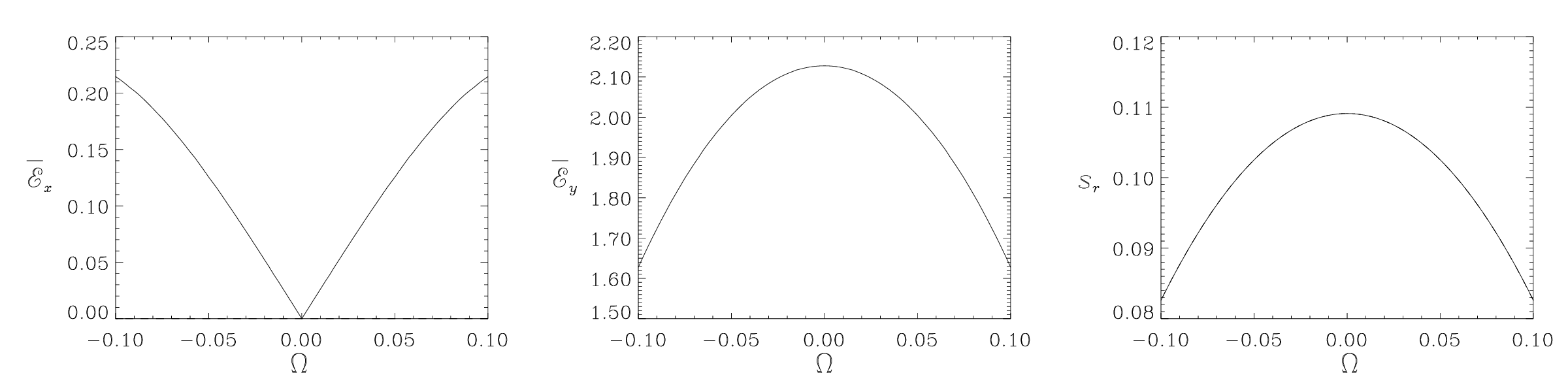}
\caption{Variation of $\barcalE_x$, $\barcalE_y$ and $s_r$ with $\Omega$. All other parameters are as in Fig.~\ref{fig:emfs}.}
\label{fig:emfsomega}
\end{figure}

\begin{figure}[htb]
\plotone{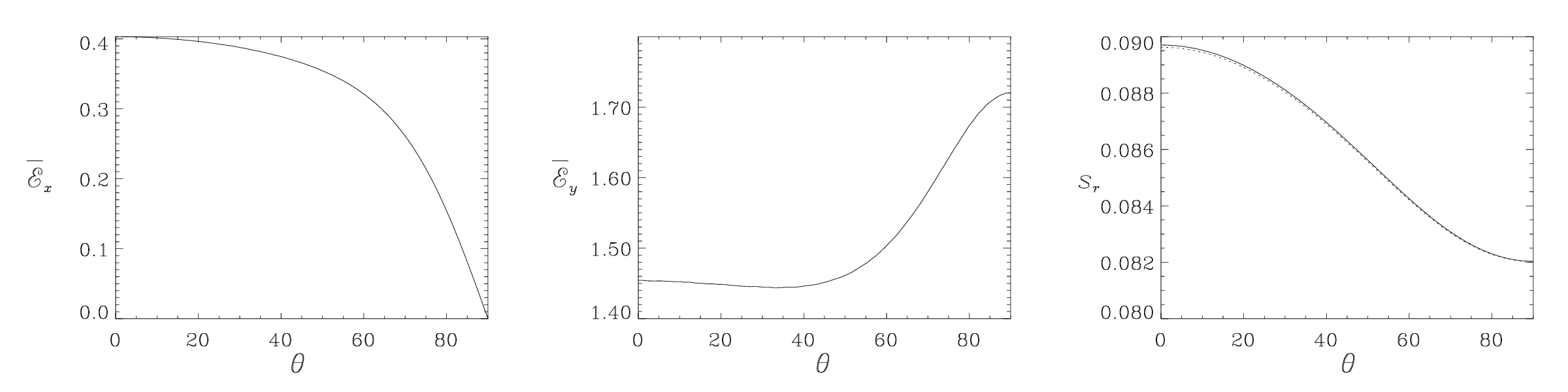}
\caption{Dependence of $\barcalE_x$, $\barcalE_y$ and $s_r$ on $\theta$. All other parameters are as in Fig.~\ref{fig:emfs}.}
\label{fig:emfscolat}
\end{figure}

\begin{figure}[htb]
\plotone{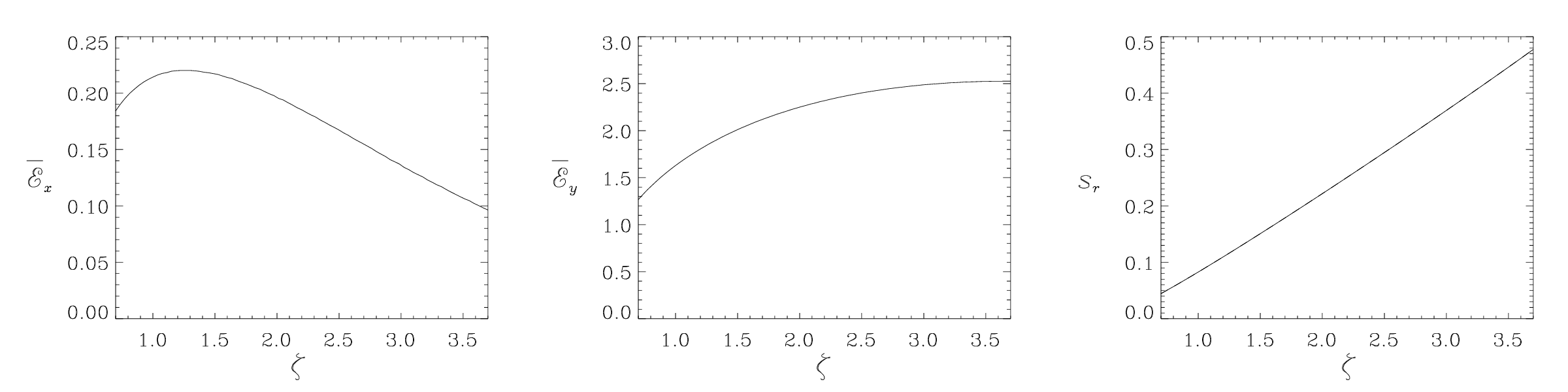}
\caption{Dependence of $\barcalE_x$, $\barcalE_y$ and $s_r$ on $\zeta$. All other parameters are as in Fig.~\ref{fig:emfs}.}
\label{fig:emfszeta}
\end{figure}

\begin{figure}[htb]
\plotone{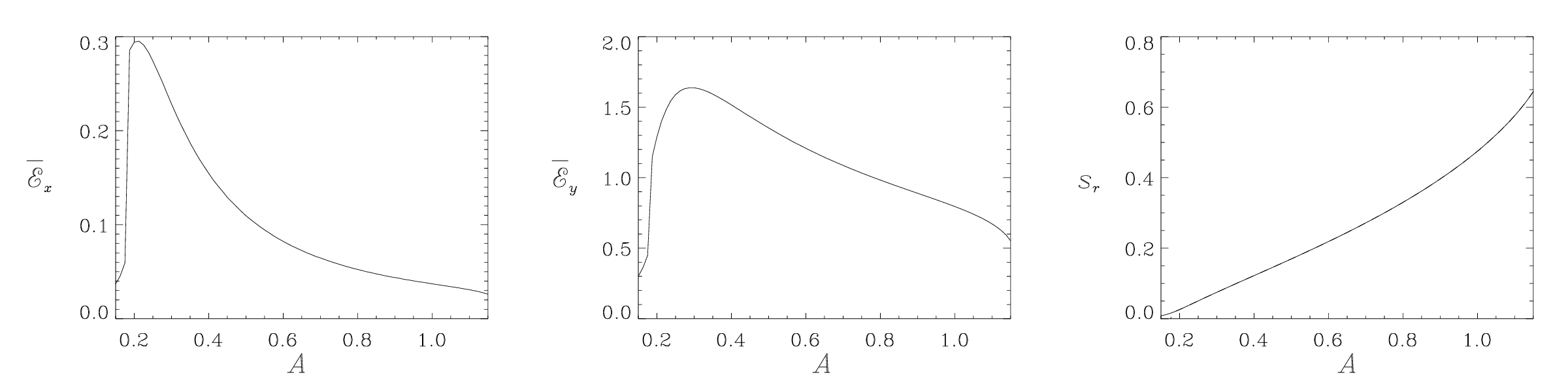}
\caption{Dependence of $\barcalE_x$, $\barcalE_y$ and $s_r$ on $A$. All other parameters are as in Fig.~\ref{fig:emfs}.}
\label{fig:emfsb0}
\end{figure}

It is also of interest to examine how the emf varies with the parameters of the problem. In order to do this and to be able to display the results in a meaningful way, we must first introduce an overall ($z$-independent) measure of the emf; with this in mind, we define $\bfbarcalE$ where $\barcalE_i=\left(\int_0^1\bfcalE_i^2 dz\right )^{1/2}$. Figures~\ref{fig:emfsomega} -- \ref{fig:emfsb0} show the variation of $\bfbarcalE$ with $\Omega$, $\theta$, $\zeta$ and $B_0$ respectively. Throughout all simulations, the dimensionless magnetic diffusivity, viscosity and gravitational field strength were kept fixed. The first two plots in each set of graphs show $\barcalE_x$ and $\barcalE_y$ as functions of the parameter under investigation. The third shows the growth rates of the fastest growing modes, optimised over $k$ and $l$, in the $k\mbox{, }l>0$ (solid line) and $k<0\mbox{, }l>0$ (dashed line) directions; as can be seen, these curves are essentially identical. Each graph is constructed using the weighting procedure defined by equation~(\ref{eq:weighted_emf}). As discussed in \S\,\ref{sec:formulation}, this provides a sensible description of those cases for which two waves travelling in opposite directions produce cancelling emfs (such as at $\theta=\pi/2$). It can be seen that this feature is reflected in our results.

Figure~\ref{fig:emfsomega} shows the variation of $\barcalE_x$, $\barcalE_y$ and growth rate $s_r$ with angular velocity $\Omega$. The symmetry considerations leading to equation~(\ref{eq:radler}) suggest that $\barcalE_x \propto \Omega$ and $\barcalE_y = \mbox{ constant}$. Although we do not expect these relations to hold exactly, since they do not take into account the boundaries in $z$, it can be seen that they are reasonably well satisfied; $\bar \calE_x$ is approximately linear in $\Omega$ (the discontinuity in the derivative of $\barcalE_x$ simply reflects the fact that $\calE_x$ is odd in $\Omega$ whereas $\barcalE_x$ is an rms quantity), whilst $\barcalE_y$ (which is even in $\Omega$) shows some variation about a significant mean value. This reflects what is expected from the discussion of symmetries in \S\,\ref{sec:formulation}, where we explained how we would expect a change in the sign of $\Omega$ to lead to a change in the sign of $\calE_x$ but not of $\calE_y$.

In the traditional formulation of mean field electrodynamics \citep[see][]{Moffatt78}, regeneration of the mean field relies on a lack of reflectional symmetry in the flow, usually characterised by the flow possessing net helicity. Field regeneration is then ascribed to the `$\alpha$-effect', which is given by the symmetric part of the $\alpha_{ij}$ tensor in expression~(\ref{eq:emf_expansion}); the antisymmetric part of $\alpha_{ij}$ is interpreted as a mean pumping velocity for the magnetic field. As discussed above, the decomposition~(\ref{eq:emf_expansion}) is not meaningful for the problem we are considering. However it is still meaningful to examine which parts of the emf depend on reflectionally symmetric parts of the physics. This is most readily achieved by changing the sign of $\Omega$ -- the elements of $\bfcalE$ that are reflectionally symmetric will change sign with $\Omega$ while the non-reflectionally symmetric parts will not. In the absence of any rotation, $\calE_x$ is zero while $\calE_y$ is at its largest in magnitude. This leads us to associate $\calE_y$ with magnetic pumping and turbulent diffusion.

Figure~\ref{fig:emfscolat} displays the dependence of $\barcalE_x$, $\barcalE_y$ and $s_r$ on colatitude $\theta$; note that the instability is more vigorous at the poles. Using only simple symmetry arguments, equation~(\ref{eq:radler}) suggests that $\barcalE_x\propto \cos\theta$ and $\barcalE_y=\mbox{ constant}$. The agreement actually is quite good at high latitudes, but rather less so for mid-to-low latitudes.

Figure~\ref{fig:emfszeta} shows the variation of $\bfbarcalE$ with the basic state magnetic field gradient $\zeta$. It can be seen that the growth rate increases approximately linearly with $\zeta$, confirming the idea that a larger field gradient leads to a more vigorous instability. The magnitude of both $\barcalE_x$ and $\barcalE_y$ increases with $\zeta$ for small values, which is to be expected as the instability sets in and becomes stronger. In general, the significance of $\barcalE_x$ reflects a balance between the vigour of the instability and the influence of rotation, as illustrated by the picture of \citet{Parker55} of rising, twisting flux loops (termed `cyclonic events' by Parker). If the instability is weak then $\barcalE_x$ is also weak; conversely if the instability is strong, the influence of rotation is diminished and $\barcalE_x$ is reduced. This feature can be seen clearly in the plot of $\barcalE_x$ versus $\zeta$.

Figure~\ref{fig:emfsb0} shows the variation of $\barcalE_x$, $\barcalE_y$ and $s_r$ with $A$, which corresponds to a change in the strength of the basic state magnetic field at the top of the layer (since $\rho (z=0)$ is kept constant). For low values of $B_0$, $\barcalE_x$ increases --- this can be thought of as an increase in the field strength allowing the instability to become stronger and therefore generate more emf. As the strength of the field increases further, the growth rate continues to increase; the influence of rotation is thus diminished and, just as in Figure~\ref{fig:emfszeta}, $\barcalE_x$ decreases in magnitude. There is a sharp increase in $\barcalE_y$ over low Alfv\'en speeds as the instability sets in, a slight decrease as the Alfv\'en speed is further increased and then a sharp drop at high Alfv\'en speeds.

\section{Sheared Magnetic Field}
\label{sec:sheared}

Since for many astrophysical dynamos it is believed that the toroidal field, generated by the differential rotation (the `$\omega$-effect' in the language of mean field electrodynamics), dominates the poloidal component, then it makes sense initially to consider the nature of the emf that results from the instability of a purely toroidal field; this is the approach adopted in \S\S\,\ref{sec:ideal},\,\ref{sec:unidirectional}. The dynamo resulting from considering only the $\omega$-effect and the mean emf resulting from a unidirectional field may be considered as the analogue of the classical $\alpha \omega$-dynamo. However, in reality, the generation of a mean $y$-component of the field will influence the instability to some degree, and it is therefore of interest to extend the analysis of \S4 so as to consider the nature of the emf arising from the instability of a sheared horizontal magnetic field. In this section, therefore, we consider a basic state magnetic field of the form $\bfbarB=\Bbar_x\mathbf{e}_{x}+\Bbar_y\mathbf{e}_{y}$, where in non-dimensional form $\bfbarB_x=(1+\zeta z)$ as before, and $\bfbarB_y= \lambda (1+\xi z)$. The extended problem thus has two additional parameters: $\xi$, the gradient of the $y$-component of the basic state magnetic field, and $\lambda$, the ratio of the $y$-component of the basic state magnetic field to the $x$-component at the top of the layer.

The influence on the dynamics of the instability of a weak $y$-component of the magnetic field is of interest in itself, with applications to the morphology of the field that may emerge from the solar tachocline. The inclusion of $\Bbar_y$ leads not only to changes in the stratification, but also to an additional restraint on the instability, since any instability must `unwind' the field. We propose to discuss the details of the instability in a separate paper; here we shall concentrate on its influence on the emfs generated. \citet{CCH89} have also considered magnetic buoyancy instability of a sheared magnetic field, but they restricted attention to the non-rotating case; in order to determine a preferred (`toroidal') direction they made the restriction to $x$-independent perturbations ($k=0$). In our system, the presence of rotation distinguishes the $x$- and $y$-directions and we may therefore consider fully three-dimensional disturbances, without ambiguity.

\begin{figure}[htb]
\plotone{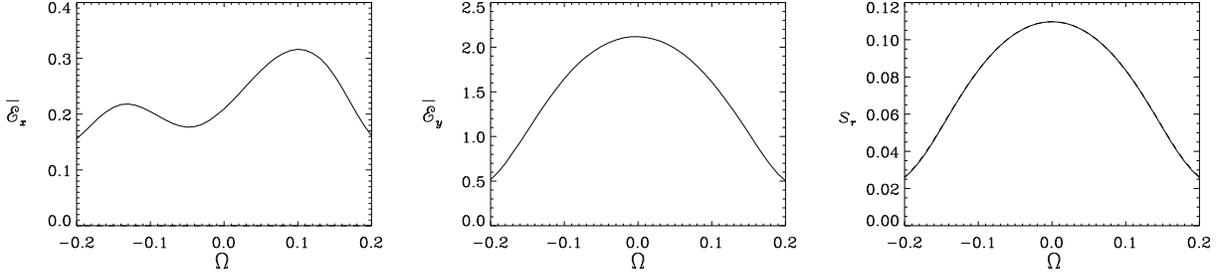}
\caption{Dependence of $\barcalE_x$, $\barcalE_y$ and $s_r$ on $\Omega$ in the case with a sheared basic state magnetic field. Parameter values are $\theta=75^{\circ}$, $A^{2}=0.1$, $\zeta=1$, $\tilde \nu = \tilde \eta = 5\times 10^{-5}$, $\chi =2$, $\lambda=0.1$ and $\xi=1$.}
\label{fig:emfs2domega}
\end{figure}

\begin{figure}[htb]
\plotone{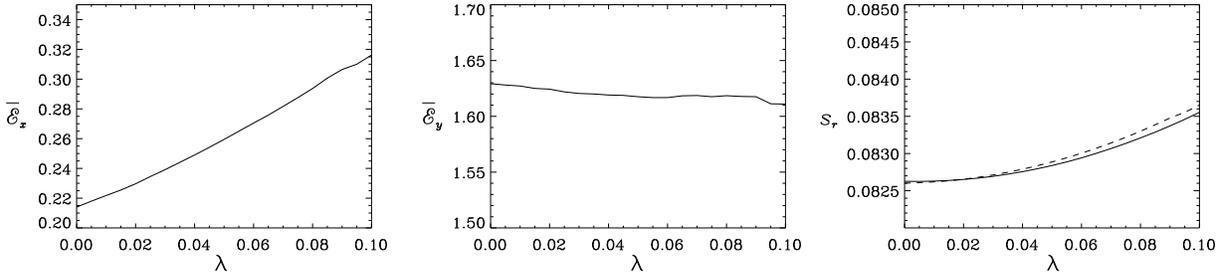}
\caption{Dependence of $\barcalE_x$, $\barcalE_y$ and $s_r$ on $\lambda$ at the top of the layer. Parameters as in Fig.~\ref{fig:emfs2domega}, with $\Omega=0.1$. In the third panel, the dashed and solid lines represent the growth rates of the two fastest-growing modes, which are then used in the weighting process.}
\label{fig:emfsby0}
\end{figure}

\begin{figure}[htb]
\plotone{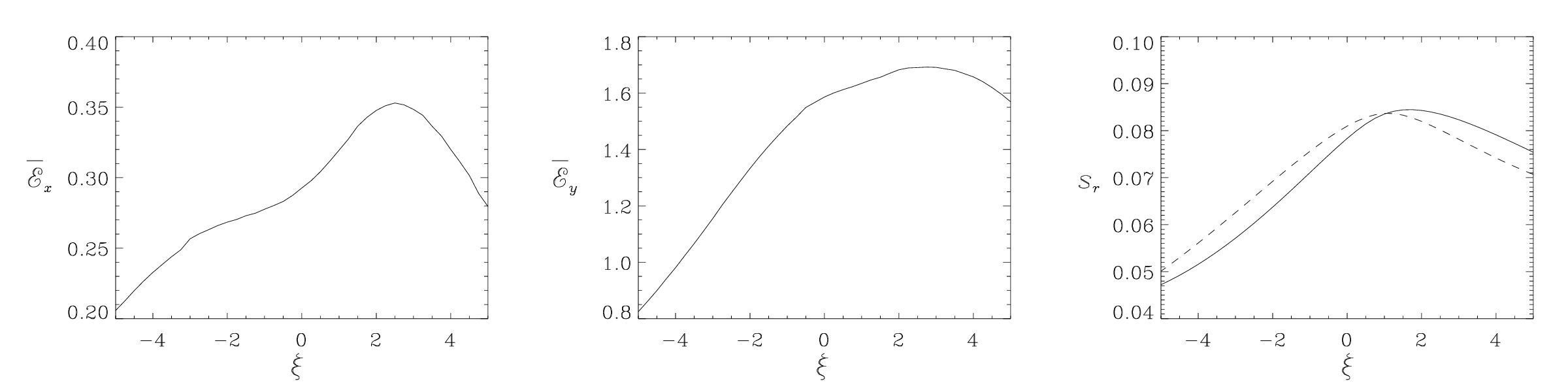}
\caption{Dependence of $\barcalE_x$, $\barcalE_y$ and $s_r$ on $\xi$. Parameters as in Figs.~\ref{fig:emfs2domega} and \ref{fig:emfsby0}.}
\label{fig:emfsxi}
\end{figure}

Figures~\ref{fig:emfs2domega} -- \ref{fig:emfsxi} illustrate the dependence of $\bfbarcalE$ on various parameters, with emfs weighted using the same prescription as in \S\,\ref{sec:unidirectional}. Unlike in \S\,\ref{sec:unidirectional}, however, the two fastest-growing modes no longer necessarily have oppositely-signed $x$-direction wave numbers. For a unidirectional magnetic field, only the angular rotation $\bfOmega$ breaks the symmetry between the $(k,l)$ and $(-k,l)$ modes. The introduction of a poloidal ingredient means that there is an additional symmetry breaking, and so there is even less reason to expect the fastest growing modes to have oppositely-signed $k$. The weighting procedure detailed in \S\,\ref{sec:formulation} is still applied so that the results can be compared directly in the case of $\Bbar_y\to 0$ with those of \S\,\ref{sec:unidirectional}. 



Figure~\ref{fig:emfs2domega} displays the variation of $\barcalE_x$, $\barcalE_y$ and $s_r$ with $\Omega$. The growth rate and both components of the emf are diminished as $|\Omega|$ is increased. The growth rates of the two fastest-growing modes are very similar, as can be seen from the fact that the solid and dotted lines in the figure are indistinguishable. Unlike in Figure~\ref{fig:emfsomega}, the addition of the poloidal component $B_y$ means that $\calE_x$ is no longer antisymmetric with respect to $\Omega$, and, although from the graph it is not discernible, $\calE_y$ is no longer symmetric in $\Omega$. With non-zero $\bfbarB_y$, the non-rotating case no longer represents a significant case of symmetry; when $\Bbar_y=0$, the case of $\Omega=0$ represents a situation in which the poloidal and toroidal directions cannot be distinguished --- the introduction of the poloidal component removes this property. Now, $\barcalE_x$ has both reflectionally symmetric and non-reflectionally symmetric parts, as does $\barcalE_y$.


Figure~\ref{fig:emfsby0} shows the variation of $\barcalE_x$, $\barcalE_y$ and $s_r$ with $\lambda$ (the ratio $\Bbar_y/\Bbar_x$ at the top of the layer). As we have chosen a case with $\xi > 0$, both the $x$- and $y$-components of $\bfbarB$ are, individually, potentially destabilising (field strength increasing with depth); increasing $\lambda$ therefore increases an already destabilising component of the field. As the strength of the poloidal field increases, the growth rate increases, as does the emf in the $x$-direction, while the emf in the $y$-direction slightly decreases in strength. The increase in the value of $\barcalE_x$ can be viewed as the instability becoming stronger as $\lambda$ is increased. As expected, the difference between the two fastest-growing modes increases with $\Bbar_y$, as is evident from the separation of the two lines in Figure~\ref{fig:emfsby0}(c). 

Figure~\ref{fig:emfsxi} shows the dependence of $\barcalE_x$, $\barcalE_y$ and $s_r$ on $\xi$, the gradient of $B_y$, for $\zeta = 1$. Both positive and negative values of $\xi$ are considered --- positive values of $\xi$ correspond to a basic state field in which both the $x$- and $y$-components can be thought of as destabilising, while negative values of $\xi$ correspond to a stabilising $B_y$-component. When $\xi<0$, both the emf and the growth rate increase as $\xi$ becomes less negative (i.e., as the poloidal field increases less with height), and then continue to increase with $\xi$ for low values of $\xi$. Eventually the emfs and growth rates all begin to decrease as $\xi$ increases further.

In their study of the magnetic buoyancy instability of a sheared magnetic field, with the crucial restriction to $x$-independent perturbations, \citet{CCH89} found that the nature of the instability was strongly influenced by the location of the height at which $B_y =0$, where the field appears locally untwisted to $x$-independent perturbations (the \textit{resonant surface}). It is therefore of interest to examine this feature in our system, although it is important to note that here we are considering fully three-dimensional instabilities, for which the notion of a unique untwisted direction is not clearly defined. We have examined the effect of varying the depth at which $\Bbar_y=0$ (henceforth called $z_0$), whilst keeping the magnetic energy of the $y$-component of the field constant. The ratio of the magnetic energies of the $y$- and $x$-components of the basic state magnetic field is $3/7$.

\begin{figure}[htb]
\plotone{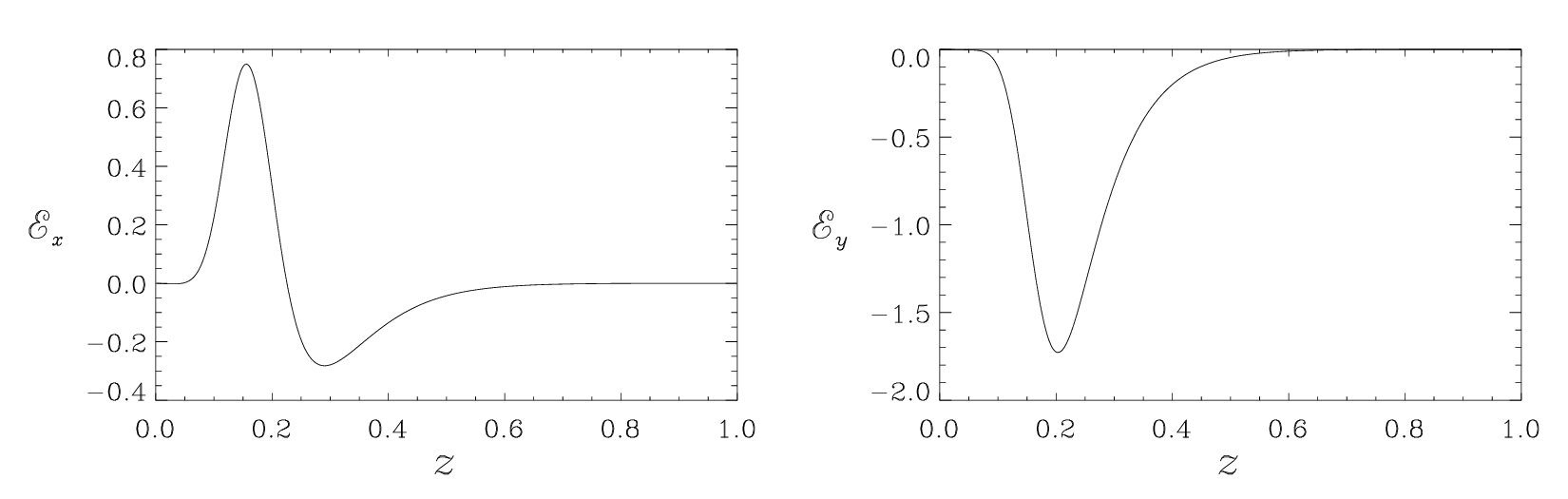}
\caption{$\calE_x$ and $\calE_y$ for the the most unstable mode in the case $z_0=0.1$. The wave numbers are $k=0.18987$ and $l=7.4219$, and the eigenvalue is $s=0.050992-0.043032i$.}
\label{fig:emfsz01}
\end{figure}

\begin{figure}[htb]
\plotone{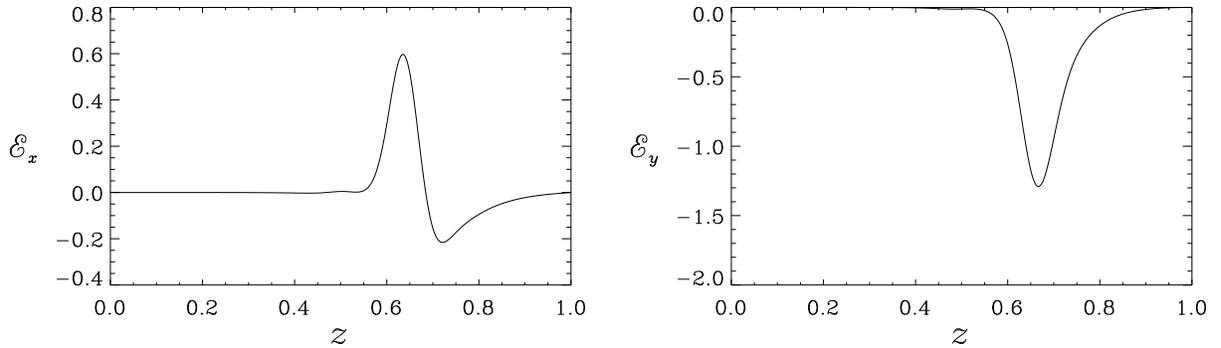}
\caption{$\calE_x$ and $\calE_y$ for the most unstable mode in the case $z_0=0.5$. The wave numbers are $k=-0.35297$ and $l=7.4875$, and the eigenvalue is $s=0.02385-0.025888i$.}
\label{fig:emfsz05}
\end{figure}

\begin{figure}[htb]
\plotone{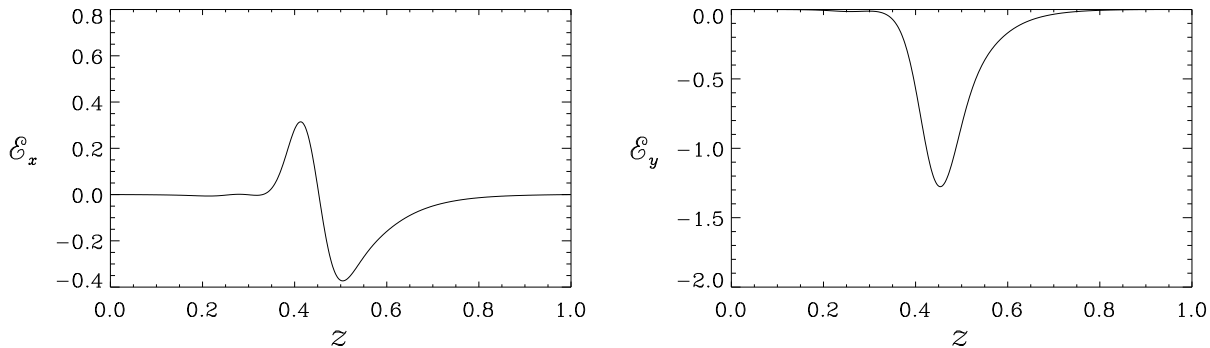}
\caption{$\calE_x$ and $\calE_y$ for the most unstable mode in the case $z_0=0.9$. The wave numbers are $k=1.6178$ and $l=6.1094$, and the eigenvalue is $s=0.021667-0.023000i$.}
\label{fig:emfsz09}
\end{figure}

Figures~\ref{fig:emfsz01}, \ref{fig:emfsz05} and \ref{fig:emfsz09} show $\calE_x$ and $\calE_y$ as functions of depth for three values of $z_0$: $z_0=0.1$ ($\Bbar_y=0$ near the top of the layer), $z_0=0.5$ ($\Bbar_y=0$ in the middle of the layer) and $z_0=0.9$ ($\Bbar_y=0$ near the bottom of the layer). Results with $z_0<0$ or $z_0>1$ are extremely similar to those with $z_0$ near the top or bottom of the layer respectively, and so are not displayed here. The quantities displayed are $\calE_x$ and $\calE_y$, calculated using the eigenfunctions of the single most unstable mode, unweighted so as to allow direct comparison with Figure~\ref{fig:emfs} (which is an example of an unweighted emf resulting from a unidirectional basic state field).

It can be seen that the components of the emfs in each case take very small values throughout much of the layer; the depths at which $\calE_x$ and $\calE_y$ are significant though depend on $z_0$. For all values of $z_0$, the emfs assume a similar shape to those seen in Figure~\ref{fig:emfsz01}. Comparing the three plots of $\calE_x$ in Figures~\ref{fig:emfsz01} -- \ref{fig:emfsz09}, we can see that as $z_0$ is increased, the first peak (with $\calE_x>0$) decreases in height, while the second (with $\calE_x<0$) is amplified. The variation of the distribution of the emfs within the layer is of interest. For a unidirectional field (Figure~\ref{fig:emfs}), $\calE_x$ and $\calE_y$ are essentially non-zero only close to the top of the layer, in a similar manner to Figure~\ref{fig:emfsz01}. In cases where $z_0$ lies either far above or far below the layer, the untwisted surface plays no role in the instability mechanism and so we expect a modified version of the unidirectional case. However, when $z_0$ is located within the layer, then the upper region of the layer has $|\Bbar_y|$ increasing with height, and so has a stabilising influence, while the lower region has $|\Bbar_y|$ decreasing with height, a destabilising influence. This suggests that the eigenfunctions should, in general, be dominant in the region $z>z_0$, i.e.\ in the least stable region, as can be seen in Figure~\ref{fig:emfsz05}. When $z_0$ is close to the bottom boundary, this is not possible, owing to the boundary conditions, and so the eigenfunctions are non-zero towards the middle of the layer, as seen in Figure~\ref{fig:emfsz09}.

\section{Conclusions and Discussion}
\label{sec:conc}
We have examined the nature of the mean electromotive force that arises from rotationally influenced magnetic buoyancy instability of a stratified magnetic field. In a standard kinematic mean field dynamo the magnetic field is amplified from an infinitesimal strength via an $\alpha$-effect that results from consideration solely of the velocity field; an instability-driven dynamo, on the other hand, is intrinsically dynamic, in that the motions responsible for driving the emf arise only from the instability of the mean magnetic field itself. Although instability-driven emfs have been studied in the context of magnetic confinement devices, they have received relatively little attention in the astrophysical literature. However, magnetically-driven instabilities arise in a number of astrophysical contexts and, for this reason, together with possible difficulties associated with the standard mean field approach, we believe it is of some interest to investigate this idea in some detail.

One of the most important points to note is that for an instability-driven emf, the standard decomposition (\ref{eq:emf_expansion}) is no longer meaningful and that it is the total emf that is of interest here. It is though still possible to decompose the emf into components that do and do not depend on a lack of reflectional symmetry in the flow. The dependence of the emf on the mean field and its gradient can be extremely complicated, as shown by expression~(\ref{eq:emfx_Gilman}), which considers the relatively simple case of an ideal MHD analysis with $\nu$ = $\eta =0$. For the more general treatment of the instability, discussed in \S\S\,\ref{sec:unidirectional},\,\ref{sec:sheared}, the stability problem and the subsequent evaluation of the emf have to be performed numerically. In order to treat the competition between modes of equal growth rates, but different emfs, it is necessary to introduce some weighting procedure, such as that described by expression (\ref{eq:weighted_emf}). We can then obtain both the latitudinal and depth dependence of the emfs, as well as exploring the dependence on parameters such as rotation.

We have shown that a net emf arises naturally from the preferred modes of magnetic buoyancy instabilities in a rotating system. Our results also make clear that the relationship between the effectiveness of the emf and the vigour of the instability is not straightforward --- this reinforces the important message that an instability \textit{per se} cannot guarantee a mean emf and hence a closing of the dynamo loop.

Our work described here has addressed only the nature of the mean emf that arises from magnetic buoyancy instability. The crucial next stage is to investigate the viability of the wider dynamo process, by incorporating such emfs into a model of the evolution of the mean field. We propose first to consider a highly simplified model, with the emfs taking prescribed functional forms of the mean field and its derivatives, based on the model in \S\,\ref{sec:ideal}, before considering an extended model in which the instability of the mean field and its evolution under the resulting emf are calculated self-consistently. This latter approach has some similarities with that adopted by \cite{TBG09}, who considered the hydrodynamic problem of the diffusion of the mean vorticity profile, through consideration of the products of linear fluctuations; a crucial difference though is that in their problem the response was wavelike, not the result of an instability. 

Even in the broadest sense, there is currently no consensus of how the solar dynamo operates. Three possible scenarios have been advanced \citep[see, for example,][]{TW07}: a distributed dynamo, in which field regeneration takes place throughout the convection zone; an interface dynamo, in which all the action is concentrated at the base of the convection zone and in the tachocline; and a flux transport dynamo, which might be envisaged as having an $\omega$-effect arising from the velocity shear in the tachocline and an $\alpha$-effect located close to the solar surface, with the two linked by some large-scale flow. There are certainly problems with all three types of dynamo. A distributed dynamo relies on a coherent $\alpha$-effect throughout the convection zone; this is not supported by theoretical arguments \citep{CH09} or numerical simulations \citep{BMT04}. A flux transport dynamo relies crucially on the migration of surface magnetic features and a large-scale flow to return the field to the base of the convection zone, essentially ignoring all the turbulent dynamics of the convection zone. An interface dynamo is more appealing in that the different aspects of the dynamo process are not widely spatially separated. Toroidal field would be amplified from the poloidal component by the differential rotation in the tachocline; the difficulty though is in closing the dynamo cycle. One possibility is from helical overshooting convection, though this would then still rely on the traditional $\alpha$-effect working at high magnetic Reynolds numbers. An emf resulting from magnetic buoyancy instability provides a natural possible alternative solution to this problem. Clearly, further, more involved, investigations are needed in order to examine the viability of this idea.

\acknowledgments
C.R.D.\ was supported by a studentship from the Science and Technologies Facilities Council. D.W.H.\ was supported by STFC and by a Royal Society Leverhulme Trust Senior Research Fellowship.

\end{document}